\documentclass[aps,prl,twocolumn,groupedaddress,showpacs,floatfix,superscriptaddress,longbibliography]{revtex4-2}
\usepackage[letterpaper,margin=1in]{geometry}
\usepackage[plainpages=false,pdfpagelabels,colorlinks=true,linkcolor=black,urlcolor=black,citecolor=black,pdftitle={Title},pdfauthor={},pdfdisplaydoctitle=true,pdfduplex=DuplexFlipLongEdge]{hyperref}
\usepackage{siunitx}
\usepackage{mhchem}
\usepackage{epsfig}
\usepackage{graphicx}
\usepackage[utf8]{inputenc}
\usepackage{amsmath,amssymb}
\usepackage[dvipsnames]{xcolor}
\usepackage[normalem]{ulem}

\renewcommand\thesection{\Alph{subsection}}


\usepackage{titlesec}
\titleformat{\section}{\normalsize\bfseries}{\thesection}{1em}{}  
\titleformat{\subsection}{\small\bfseries}{\thesection}{1em}{}
\titlespacing*{\section}{0pt}{*2}{*0.5}
\titlespacing*{\subsection}{0pt}{*0.5}{*0.2}

\begin{document}

\title{Hydrogen Bond Strength Dictates the Rate-Limiting Steps of Diffusion in Proton-Conducting Perovskites: A Critical Length Perspective}
\author{Hang Ma}
\affiliation{School of Physics and Astronomy, and Key Laboratory of Multiscale Spin Physics (Ministry of Education), Beijing Normal University, Beijing 100875, China\\}
\author{Jiajun Linghu*}
\affiliation{ Department of Applied Physics, Chang'an University, Xi'an 710064, China\\}
\author{Nannan Han}
\affiliation{State Key Laboratory of Flexible Electronics (LOFE) \& Institute of Flexible Electronics (IFE), Northwestern Polytechnical University, 127 West Youyi Road, Xi'an 710072, China\\}
\author{Peng Feng}
\affiliation{State Key Laboratory of Flexible Electronics (LOFE) \& Institute of Flexible Electronics (IFE), Northwestern Polytechnical University, 127 West Youyi Road, Xi'an 710072, China\\}
\author{Yuling Zhuo}
\affiliation{State Key Laboratory of Flexible Electronics (LOFE) \& Institute of Flexible Electronics (IFE), Northwestern Polytechnical University, 127 West Youyi Road, Xi'an 710072, China\\}
\author{Ying Liang*}
\affiliation{College of Physics, Hebei Normal University, Shijiazhuang 050024, China}
\affiliation{School of Physics and Astronomy, and Key Laboratory of Multiscale Spin Physics (Ministry of Education), Beijing Normal University, Beijing 100875, China\\}
\author{Kunde Yang*}
\affiliation{Ocean Institute of Northwestern Polytechnical University, Taicang, Jiangsu 215400, China\\}
\author{Tianxing Ma*}
\affiliation{School of Physics and Astronomy, and Key Laboratory of Multiscale Spin Physics (Ministry of Education), Beijing Normal University, Beijing 100875, China\\}
\author{Zhi-Peng Li*}
\affiliation{State Key Laboratory of Flexible Electronics (LOFE) \& Institute of Flexible Electronics (IFE), Northwestern Polytechnical University, 127 West Youyi Road, Xi'an 710072, China\\}

\begin{abstract}
Identifying the rate-limiting step of proton migration in proton-conducting oxides is essential for assessing and regulating proton conductivity. Proton migration based on the Grotthuss mechanism involves both proton rotation and proton transfer, with the latter typically regarded as the rate-limiting step. However, a universal criterion for identifying the rate-limiting step remains to be established. Here, we perform a quantitative decomposition of the rotation and transfer barriers, revealing that the hydrogen bond to the acceptor oxygen dictates their energy barrier difference via the O$_i$-B-O$_f$ bending mechanism. Based on the energy difference associated with a one-order-of-magnitude variation in residence time, we propose the hydrogen bond length criterion for identifying the rate-limiting step across operating temperatures. Taking the 500 K criterion as an upper limit, when the hydrogen-bond length of systems falls below 2.05~\AA, proton rotation becomes competitive with transfer. Applied to a wider range of perovskite materials, this criterion predicts comparable rotation and transfer rates in cubic structures with small lattice constants, low-valent B-site doped systems with moderate ionic radii, and distorted orthorhombic structures. Our findings provide an atomic-scale insight into the proton migration mechanisms in perovskites, and offer practical guidance for optimizing and designing advanced proton-conducting electrolytes.
\end{abstract}

\date{Version 16.0 -- \today}

\maketitle
\noindent
\section{1. INTRODUCTION}
Proton conduction is pivotal in various energy technologies particularly in solid oxide fuel cells (SOFCs)\cite{ABE201915072,XU2022115175,BICER20203670,TIMURKUTLUK20161101,doi:10.1021/acsami.3c09025,ding2020self} and solid oxide electrolysis cell (SOEC)\cite{ni2008technological,bi2014steam,lei2019progress}, which represent next-generation energy conversion devices. In solid oxides, proton diffusion primarily follows the Grotthuss mechanism, which consists of two fundamental steps: proton transfer between adjacent oxygen ions and the proton rotation around a single oxygen ion site\cite{AGMON1995456,KREUER2000149,PhysRevLett.103.238302,popov2023search}. The long-range proton diffusion is limited by the slowest step\cite{BJORKETUN20053035}. Identifying the rate-limiting step is therefore crucial for assessing and optimizing proton conductivity\cite{article,pan2023exploring,PhysRevB.87.104303}. Early experimental and theoretical studies have shown that in a considerable number of perovskite materials( SrZrO$_3$,BaCeO$_3$, BaTiO$_3$, et al.), the proton transfer is often the rate-limiting step with a higher energy barrier than that of proton rotation\cite{PIONKE1997497,MATZKE1996621,KREUER1995157,MUNCH1996647,Munch01041999,hempelmann1998muon,kreuer1998proton,munch1997quantum}.  Consequently, doping lower-valence (e.g., +3) ion on the B-site is an effective strategy to reduce the energy barrier\cite{draber2020nanoscale,HE2022111007}. However, rotation is not always one order of magnitude faster than transfer, as exemplified by orthorhombic BaCe$_{1-x}$Gd$_x$O$_{3-x/2}$\cite{PhysRevB.87.104303} and closely packed SrCeO$_3$\cite{Munch01041999}, which proton rotation exhibits comparable or even dominant barriers. And A-site doping with low-valence (e.g., +1) ions can effectively enhance the proton conductivity\cite{rfs5-bnwt,PhysRevB.76.054307,https://doi.org/10.1111/jace.14839}. This contrast highlights a key issue: despite the significance of distinguishing the rate-limiting step, a universal and direct criterion has yet to be established.

\begin{figure*}[htbp]
\begin{minipage}{0.48\linewidth}
  \centering
  \includegraphics[width=\linewidth]{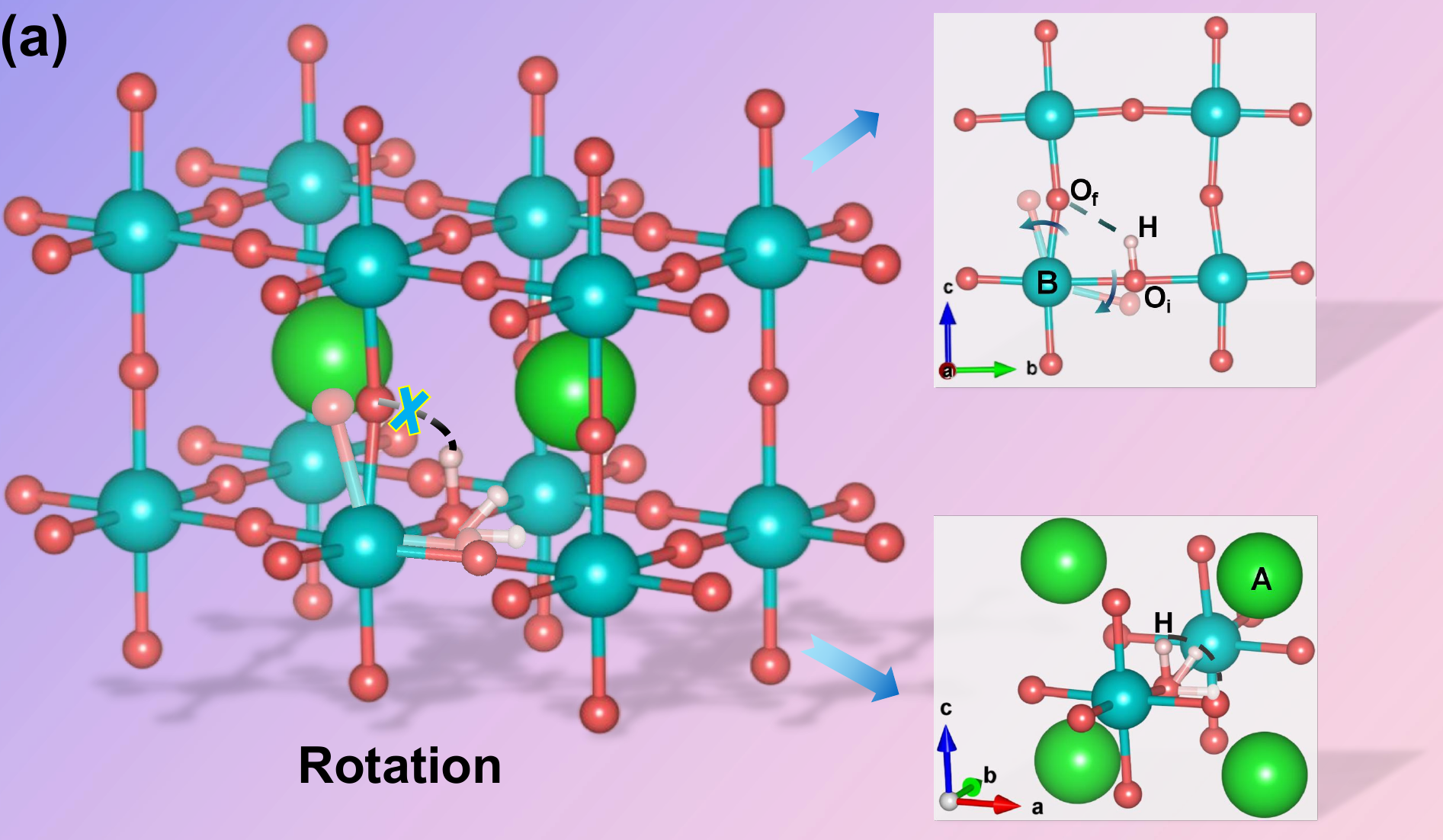}
\end{minipage}
\hfill
\begin{minipage}{0.48\linewidth}
  \centering
  \includegraphics[width=\linewidth]{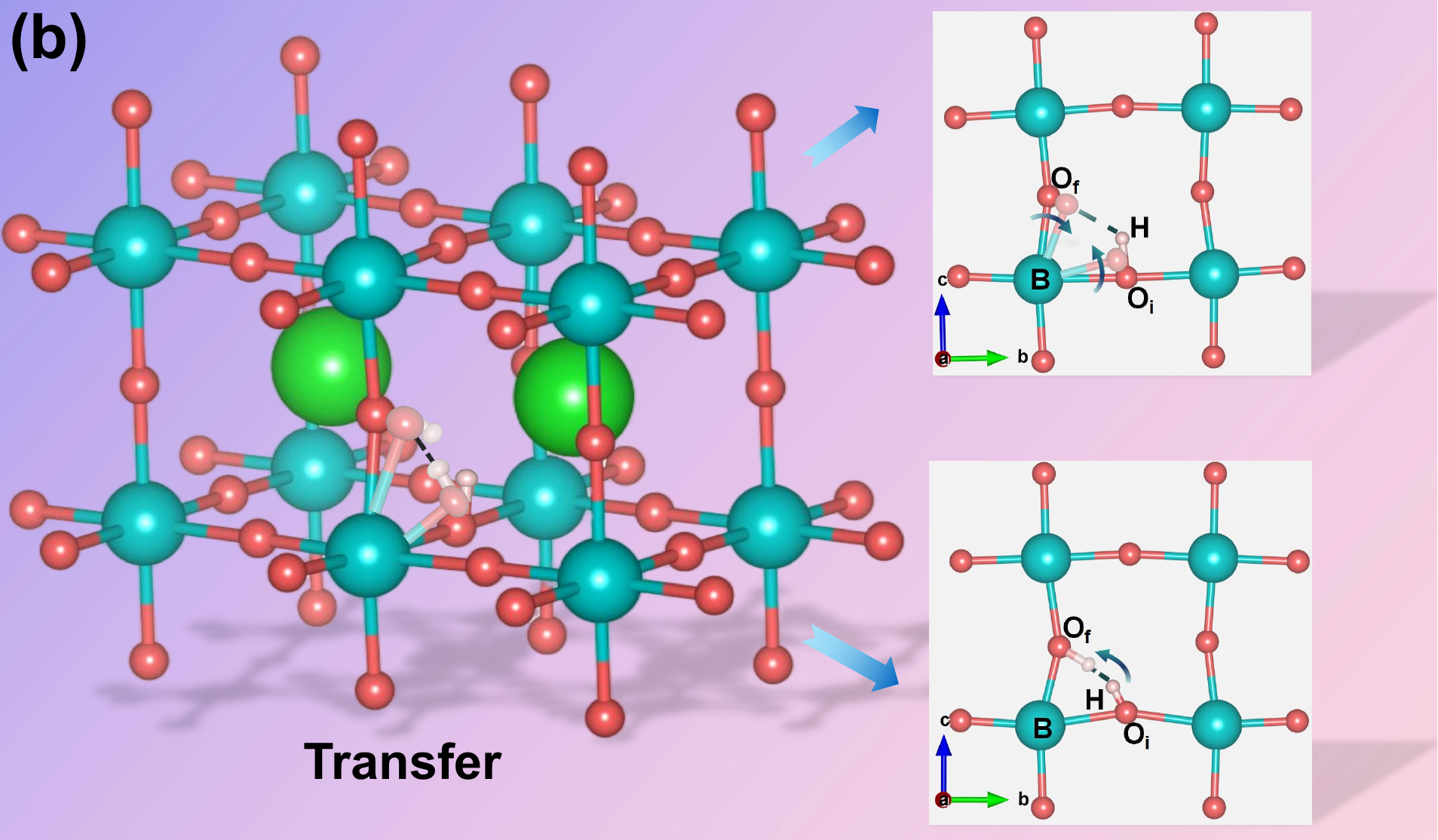}
\end{minipage}
\caption{Elementary processes of proton rotation (a) and transfer (b), respectively. 
(a) The upper panel is the schematic of outward O$_i$-B-O$_f$ bending during proton rotation. The lower panel is the reorientation of the proton around the A-site ion. (b) The upper schematic is the inward O$_i$-B-O$_f$ bending during proton transfer. The lower schematic is the proton jumping. }
\label{Fig1}
\end{figure*}

\begin{figure*}[htbp]
\centering
\includegraphics[scale=0.65]{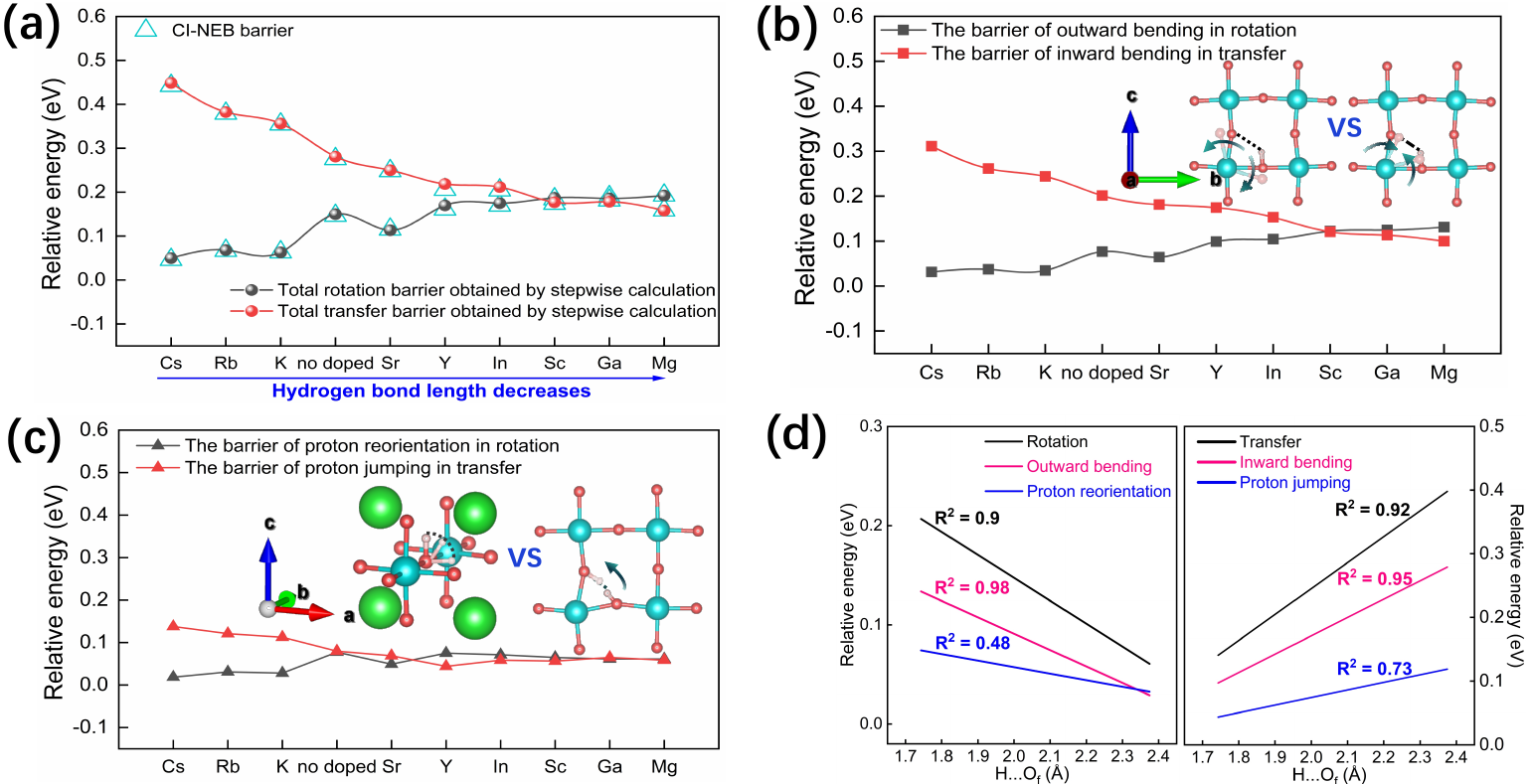}
\caption{(a) A comparison between the direct CI-NEB barriers (cyan hollow triangles) and the total barriers calculated by elementary processes shown in Fig.~\ref{Fig1}(a)(b) (spheres) in BaHfO$_3$, its A-site (M$_a$=Sr$^{2+}$,K$^{+}$,Rb$^{+}$,Cs$^{+}$), and B-site (M$_b$=Mg$^{2+}$,Ga$^{3+}$,Sc$^{3+}$,In$^{3+}$,Y$^{3+}$)doped systems. (b) The O$_i$-B-O$_f$ bending-component barriers for the rotation (black symbols) and transfer steps (red symbols). (c) The barriers of proton reorientation during rotation (black symbols) and jumping during transfer (red symbols). (d) Linear fits of the barriers for rotation and transfer, as well as their decomposed steps, as a function of the hydrogen-bond length.} 
\label{Fig2}
\end{figure*}

Early studies by Kreuer \textit{et al.} qualitatively highlighted the important role of hydrogen bonding in proton rotation and transfer\cite{kreuer1996proton,munch1997quantum,Munch01041999,KREUER1995157}. Subsequently, they also pointed out that the proton transfer barrier is strongly influenced by the repulsive interaction between the proton and the B-site cation\cite{MUNCH2000183}. More recent studies on proton diffusion in solid oxides have identified multiple factors governing migration barriers\cite{D2TA08664F,doi:10.1021/acs.chemmater.1c02432,D4EE01219D,10.1063/1.3122984,PhysRevLett.102.075506,islam2022first,chung2026flexibility}, including the hydrogen-bond length between the proton and the acceptor oxygen, the proton-A/B-site cation repulsive interaction, the O-H stretching frequency, and the metal-oxygen coordination. These factors are hierarchically important and mutually coupled. To identify a common dominant factor governing both rotation and transfer barriers—and thereby explain the large rotation-transfer barrier differences observed in many systems—we quantitatively analyze the proton-lattice coupled mechanism\cite{du2020cooperative,braun2017experimental} and isolate the primary contribution to the barrier.

BaHfO$_3$ is selected as a model system because its tolerance factor is close to unity, ensuring the structural stability of an ideal cubic perovskite. Moreover, under some doping conditions, it exhibits higher stability and proton conductivity compared to the conventional proton-conducting electrolyte BaZrO$_3$\cite{https://doi.org/10.1111/jace.15946,KANG2017738}. Using first-principles calculations on undoped and A/B-site doped BaHfO$_3$ system, the origins of energy barriers for proton rotation and transfer are quantified. We identify the hydrogen bond as the dominant factor governing the rotation-transfer energy difference, making the slower step becomes rate-limiting when the barrier difference yields a one-order-of-magnitude contrast in residence time. Accordingly, we establish a hydrogen bond length criterion for identifying the rate-limiting step at different temperatures. An upper hydrogen bond length threshold (2.05~\AA), along with the structural conditions are further provided to estimate the dominance of proton rotation versus transfer. These findings provide insight into proton-lattice coupled diffusion mechanism\cite{SAMGIN2000291,PhysRevLett.102.075506,du2020cooperative,braun2017experimental}. The targeted doping strategies we proposed based on the identified rate-limiting step aid the design of high-conductivity perovskite materials.

\noindent
\section{2. COMPUTATIONAL METHODS}
{\textbf{Basic calculation settings.} All calculations were performed using the Vienna Ab Initio Simulation Package (VASP)\cite{KRESSE199615,PhysRevB.59.1758} with the framework of density functional theory (DFT). The  projector augmented wave (PAW) method\cite{PhysRevB.54.11169} was employed to describe the ion-electron interaction. The exchange-correlation functional was treated with the Perdew-Burke-Ernzerhof (PBE) formulation of the generalized gradient approximation (GGA)\cite{PhysRevLett.77.3865}, which is appropriate for this study as it focused on relative energy difference and barriers rather than electronic properties such as band gap\cite{inorganics13040100}. A plane wave basis set with a cutoff energy of 520 eV was used. Atomic relaxation was conducted using the conjugate gradient algorithm until the maximum force on every atom was less than 0.01~eV/\AA. For $k$-point sampling, we adopted the Monkhorst-Pack scheme\cite{PhysRevB.13.5188} with a grid spacing of 0.028 \AA$^{-1}$. 

{\textbf{CI-NEB calculation settings.} Proton migration barriers were computed using the climbing image nudged elastic band (CI-NEB) method\cite{10.1063/1.1329672}. For the initial and final states, full relaxation with ISIF = 3 was performed to ensure physically realistic hydrogen bond lengths. For the intermediate images, the Quick-Min algorithm was used with fixed lattice constants (ISIF = 2)(See the Supplementary Information(SM) for details). Cubic unit cells containing one A-site atom, one B-site atom, and three oxygen atoms (BaHfO$_3$, A/B-site doped BaHfO$_3$, BaZrO$_3$, B-site doped BaZrO$_3$, BaNbO$_3$, BaTiO$_3$, SrTiO$_3$, and SrNiO$_3$) were expanded into $2\times2\times2$ supercells for CI-NEB calculations. For the A/B-site doped systems, the proton concentration is fixed to be equal to the dopant concentration (12.5\%). This is because lattice distortions induced by different proton concentrations affect both rotation and transfer barriers in a similar manner, and thus do not alter the barrier difference that is the focus of this work. Orthorhombic unit cells containing four A-site atoms, four B-site atoms, and twelve oxygen atoms (BaCeO$_3$, SrNbO$_3$, SrZrO$_3$, LaScO$_3$, and SrCeO$_3$) were treated using $1\times2\times1$ supercells. Hubbard U values and magnetic moments recommended by the Materials Project were used for strongly correlated systems, except in two cases: in SrNiO$_3$, a magnetic moment of 1.36~$\mu_B$ and a U value of 6.2~eV were applied to Ni 3d orbitals\cite{HASAN2022105920}; in Gd-doped BaZrO$_3$, a specific pseudopotential was used to freeze the 4f electrons into the core\cite{C3TA12870A} since the 4f shell of the Gd atom is the same as that of the Gd$^{3+}$ ion, omitting the need for additional U or magnetic moment settings.

{\textbf{Stepwise computational approach for rotation and transfer.} We calculated the energy barriers for the elementary steps underlying the rotation and transfer processes in undoped, A-site doped and B-site doped BaHfO$_3$ systems using density functional theory. For proton rotation, we calculated the energy barriers of outward O$_i$-B-O$_f$  bending motion and the reorientation of proton. For proton transfer, we calculated the energy barriers of inward O$_i$-B-O$_f$ bending motion and proton jumping. By identifying a series of intermediate-angle structures (including saddle state) from the initial state to the saddle state, while fixing the positions of the moving ions O$_i$, B and O$_f$, we performed structure optimization for each intermediate-angle structure, and the position of proton also should be relaxed. The relative energy of the O$_i$-B-O$_f$  bending motion can be obtained. When the O$_i$-B-O$_f$ angle is same to the angle of the saddle state, fixing the O$_i$-B-O$_f$ angle, the energy barriers of proton rotation and transfer is calculated by CI-NEB, which obtained the energy barrier for the reorientation of proton or the proton jumping. 
 
{\textbf{Machine-learning molecular dynamics simulations.} To dynamically simulate the proton diffusion, machine-learning molecular dynamics simulations (MLMD) were performed on all selected systems. For each case, we performed approximately 100 ps of MLMD simulations with a time step of 0.333 fs (Fig.~\ref{Fig5}(b), where MSD-t shows near-linear), using the NVT ensemble with a Nosé-Hoover thermostat\cite{PhysRevB.33.8822,10.1063/1.2130390}. MD sampling was carried out at the $\Gamma$ point only. The first 33 ps were run in an on-the-fly learning mode. The ML potential obtained from the DFT training of the first 33 ps was then used to perform 10000~MD steps, from which 50 structures were uniformly selected for additional DFT single-point calculations. The final atomic forces obtained from DFT show a root-mean-square error of less than 0.1 eV/\AA\ compared with MD. This confirms the accuracy of the ML potential, which was then used for the remaining 67 ps of MD simulations.

\noindent
\section{3. RESULTS AND DISCUSSION}
{\textbf{The hydrogen bond dictates the difference in rotation-transfer energy barriers.}} The structural reliability of our computational model is confirmed by the close agreement between the optimized lattice constant of BaHfO$_3$ (4.20~\AA) and previously reported theoretical (4.17~\AA) and experimental (4.209~\AA) values \cite{10.1063/1.5010969}. According to the proton-lattice coupling mechanism during proton diffusion in perovskites\cite{KREUER2000149,SAMGIN2000291,PhysRevLett.102.075506,MUNCH2000183,du2020cooperative,JING2020227327,kreuer1996proton}, we devided the proton rotation into two elementary processes: the outward bending of O$_i$-B-O$_f$ and the susequent reorientation of proton(O$_i$ is the donor oxygen covalently bonded to the proton, and O$_f$ is the acceptor oxygen forming hydrogen bond with proton), showing in Fig.~\ref{Fig1}(a), respectively. Similarly, Fig.~\ref{Fig1}(b) illustrate two processes of proton transfer: inward bending of O$_i$-B-O$_f$ and jumping of the proton between oxygen atoms. The energy cost for each process is calculated separately: computational details are described in the Methods section, and with the results in Fig.~\ref{Fig2}(a) and Section 2 of the SM. For both proton rotation and transfer in pristine, A-site doped and B-site doped BaHfO$_3$, the CI-NEB energy barrier agrees well with the sum of energy cost of two elementary processes, demonstrating the nature of the proton-lattice coupling mechanism for both proton rotation and transfer. Notably, from Fig.~\ref{Fig2}(b)(c), the energy cost of the outward and inward bending accounts for the major portion of the total rotation and transfer barriers, respectively, in agreement with the results reported by Jing \textit{et al.}\cite{JING2020227327} and M. F. Hoedl \textit{et al.}\cite{D2TA08664F}. Furthermore, by comparing Fig.~\ref{Fig2}(a-c), we find that the difference in the bending barriers also constitutes the dominant contribution to the overall difference between rotation and transfer barriers, indicating that this process is the key factor governing rotation-transfer barrier difference.

As shown in Fig.~\ref{Fig2}(a), shorter hydrogen bonds lead to lower transfer barriers but higher rotation barriers.  Kreuer attributed this facilitation of proton transfer by short hydrogen bonds to the overlap of the electronic shells of the donor and acceptor oxygens at short distances, such that the proton remains electronically screened during transfer and does not pass through a ``bare proton" state\cite{kreuer1996proton}. However, Fig.~\ref{Fig2}(d) shows that the O$_i$-B-O$_f$ bending process exhibits the strongest linear correlation with the hydrogen-bond length. Combined with the fact that this lattice distortion dominates the total rotation and transfer barriers, we propose the following interpretation. For longer hydrogen bonds, the inward O$_i$-B-O$_f$ bending required to form the strong hydrogen bond at the transition state entails a higher energy cost, resulting in a larger transfer barrier. In contrast, outward O$_i$-B-O$_f$ bending, which breaks the hydrogen bond, is energetically easier, leading to a lower rotation barrier. This opposite effect on rotation and transfer makes the hydrogen-bond length the key factor governing the rotation-transfer barrier difference, consistent with our finding that the difference in the O$_i$-B-O$_f$ bending barriers constitutes the dominant part of the total barrier difference. It should be emphasized that, since the hydrogen bonds between the proton and O$_f$ in our studied ABO$_3$-type perovskites are of moderate strength and essentially electrostatic\cite{Steiner2002}, we use the bond length to indicate hydrogen bond strength.

Pristine BaHfO$_3$ exhibits a hydrogen bond length of 2.15 \AA, corresponding to a rotation-transfer barrier difference $\Delta E$($E_{rot}-E_{trans}$) of -0.13 eV. To access a wider range of hydrogen-bond lengths, we introduce A/B-site dopants with different valence states (M$_a$=Sr$^{2+}$,K$^{+}$,Rb$^{+}$,Cs$^{+}$,M$_b$=Mg$^{2+}$,Ga$^{3+}$,Sc$^{3+}$,In$^{3+}$\\,Y$^{3+}$). This is motivated by the fact that the net attractive interaction between the dopant ions and the proton differs from that of the host ions, leading to longer hydrogen bonds for A-site monovalent dopants (2.31–2.38 \AA) and shorter ones for B-site trivalent dopants (1.74–1.97 \AA) (see Fig.~S5 and Fig.~S6 in Section 3 of the SM). The results show that A-site-doped systems yield more negative $\Delta E$ values ($<-0.13~eV$), while B-site-doped systems show less negative or positive $\Delta E$ ($>-0.13~eV$). In particularly for Mg$^{2+}$,Ga$^{3+}$,Sc$^{3+}$ doped systems, their energy barrier of rotation is even greater than the transfer. As shown in Fig.~\ref{Fig2}(a), before the inversion point (between In$^{3+}$ and Sc$^{3+}$ doping), shorter hydrogen bonds correspond to a reduced magnitude of the rotation-transfer barrier difference; beyond this point, $\Delta E$ becomes positive and increases with decreasing hydrogen-bond length. In other words, longer (weaker) hydrogen bonds lead to a larger barrier difference, whereas shorter (stronger) hydrogen bonds reduce it. Consequently, for most proton-conducting oxides\cite{PIONKE1997497,MATZKE1996621,KREUER1995157,MUNCH1996647,Munch01041999,hempelmann1998muon,kreuer1998proton,munch1997quantum,Munch01041999,PhysRevB.82.014103}, the transfer barrier significantly exceeds the rotation barrier, making proton transfer the rate-limiting step, which can be attributed to relatively long initial hydrogen bonds. To further determine the critical hydrogen-bond length at which transfer becomes rate-limiting, we next perform a quantitative classification. 

\begin{figure*}[htbp]
\includegraphics[scale=0.53]{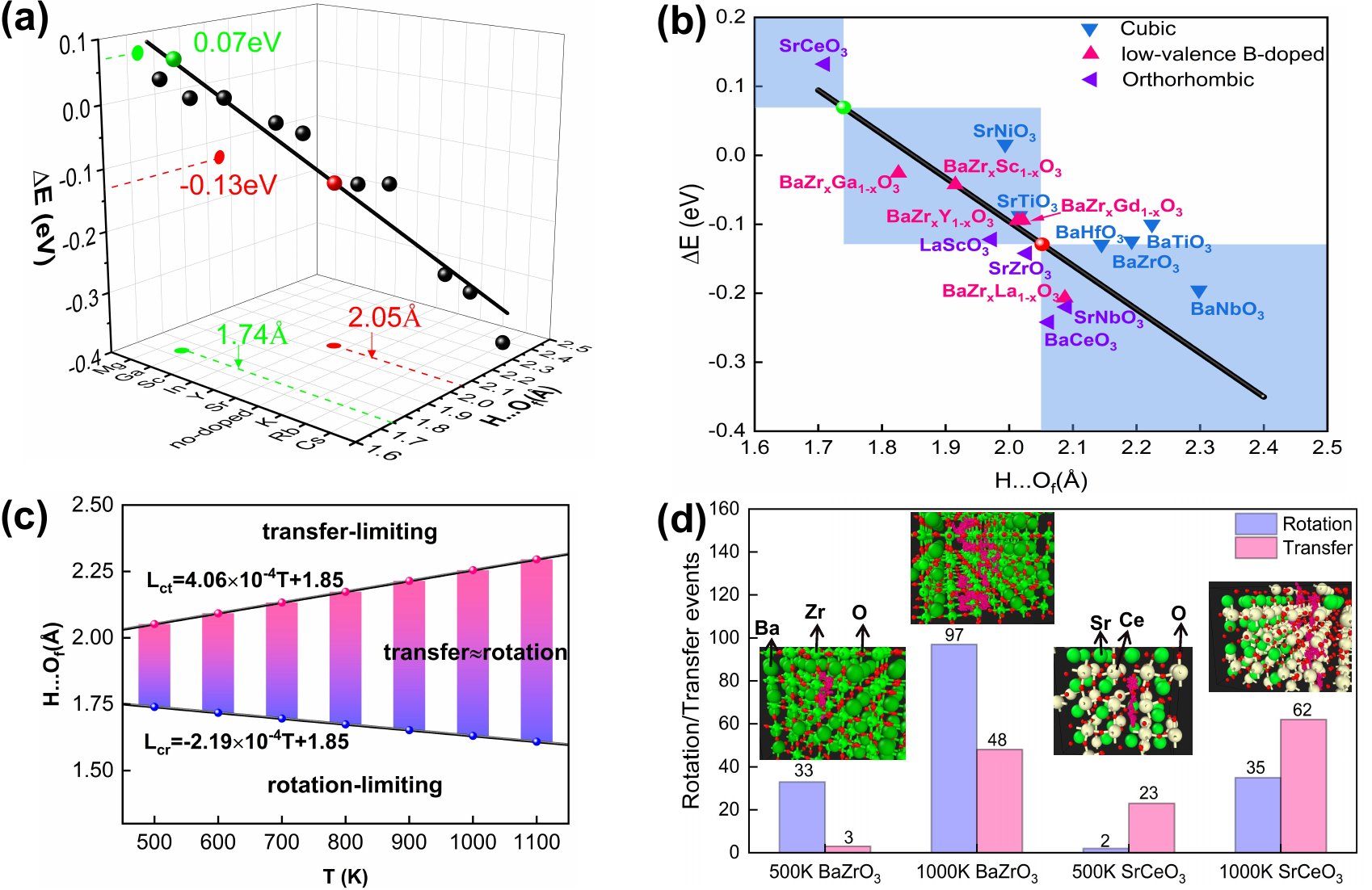}
\caption{\label{Fig3} (a) Linear interpolation fitting of H...O$_f$ and $\Delta E$ for the model system BaHfO$_3$ and its A/B-site-doped variants. Black spheres represent the original data points used in the fit, while the red and green spheres indicate the two critical points. (b) Generality test of the fitted trend in (a) by calculating the H...O$_f$ and $\Delta E$ for various perovskite materials (Detailed values and computational settings are provided in the section 5 of SM and Method). The black line represents the projection of the fitting line from panel (a) onto the H...O$_f$-$\Delta E$ plane. (c) The temperature dependence of the critical hydrogen bond lengths by linear fitting over 500-1100~K. L$_{ct}$ and L$_{cr}$ represent the critical hydrogen bond lengths for transfer-limiting and rotation-limiting. (d) The statistics of rotation and transfer events were obtained from 100 ps MD simulations of BaZrO$_3$ and SrCeO$_3$ at 500 K and 1000 K. The pink lines in the insets represent the proton trajectories.}
\end{figure*}

{\textbf{The ground-state hydrogen bond length criteria for identifying rate-limiting steps at different temperatures.}} Based on the transition state theory\cite{RevModPhys.62.251}, the residence time $t$ for proton rotation or transfer processes is defined as:
\begin{equation}
t=\Gamma^{-1}=\nu_0^{-1}e^{\frac{E}{kT}}
\label{Eq1}
\end{equation}  
where $\Gamma$ is the probability of rotation or transfer, $k$ is the Boltzmann constant, $T$ is the temperature, $\nu_0$ represents the attempt frequency and $E$ is the ground-state energy barrier. The attempt frequencies are set to $\nu_0^{rot} = 1500~cm^{-1}$ for rotation and  $\nu_0^{trans} = 3000~cm^{-1}$ for transfer, following established practice\cite{PhysRevB.87.104303}. At an operating temperature of 500 K, the energy difference $\Delta E$ of -0.13~eV between proton rotation and transfer is able to create an order-of-magnitude separation in $t$, making the transfer as the rate-limiting step of proton diffusion\cite{MUNCH1996647,Munch01041999,munch1997quantum,PhysRevB.87.104303}. Given the hydrogen bond length as the dominant factor of $\Delta E$, we establish a threshold for hydrogen bond length at which difference in $t$ between proton rotation and transfer is less than an order-of-magnitude. A linear interpolation fitting was performed for the model system BaHfO$_3$ and its A/B-site-doped variants to figure out the relationship between the hydrogen-bond lengths H...O$_f$ and $\Delta E$. Two critical points can be identified in Fig. \ref{Fig3}(a): (i) the red point with hydrogen bond length of 2.05~\AA, which corresponds to $\Delta E=-0.13~eV$, resulting in the residence time of proton transfer being an order of magnitude longer than that of rotation; (ii) the green point with hydrogen bond length of 1.74~\AA, which corresponds to $\Delta E=0.07~eV$, where the residence time of proton rotation exceeds that of transfer by an order of magnitude. These two critical points divide the entire regime into three distinct zones in Fig. \ref{Fig3}(b): the blue area in the bottom right corner with H...O$_f$ $>$ 2.05~\AA, where transfer is rate-limiting; the intermediate blue zone with 1.74~\AA $<$ H...O$_f$ $<$ 2.05~\AA\ possessing comparable transfer and rotation rates; and the blue region in the top left corner with H...O$_f$ $<$ 1.74~\AA\ dominated by rotation-limited dynamics.

We emphasize that these threshold barriers ( $\Delta E=-0.13~eV/0.07~eV$) were calculated at T = 500~K. In practice, the barrier thresholds for a one-order-of-magnitude difference in $t$ vary with temperature according to Eq.~\eqref{Eq1}, which shifts the value of the corresponding critical hydrogen bond lengths. As Fig.~\ref{Fig3}(c) shows, with increasing temperature, the critical hydrogen bond length for the transfer-limiting increases linearly (slope = $4.06\times10^{-4}$~\AA/K), while the rotation-limiting decreases (slope = $-2.19\times10^{-4}$~\AA/K). This indicates that the region with comparable proton transfer and rotation broadens at higher temperatures, which is also illustrated in Fig.~S7 of Section~4 in SM that the transfer-rotation equivalent region at higher temperatures fully encompasses that at 500 K. Therefore, the criterion obtained at 500 K is the upper limit---materials with hydrogen bond length $<$2.05~\AA\ will certainly lead to comparable proton rotation and transfer under practical operating temperatures (573-1073~K)\cite{hossain2021review}, where this region is even broader.

In order to broaden the applicability of our critical hydrogen bonds, we calculated representative perovskite materials exhibiting diverse structural features. The results (Fig. \ref{Fig3}(b)) follow an overall linear trend with a slope of -0.64, and the data points fall into distinct rotation-transfer regimes separated by the critical hydrogen-bond length. The observed generality arises from the fact that the fitted relationship captures the dominant factor governing the barrier, while additional chemical complexities in different perovskite systems act as secondary effects and do not introduce significant deviations from the overall hydrogen-bond length-barrier difference correlation. Based on the distribution of these systems in the H...O$_f$-$\Delta E$ diagram, we identify  three structural scenarios that favor comparable rate of proton transfer and rotation: (i) cubic systems with lattice constants $\leq$ 3.95~\AA, (ii) low-valent B-site doped systems with larger ionic radius mismatches $\Delta r < 0.312$~\AA, and (iii) orthorhombic systems with the tolerance factor\cite{doi:10.1126/sciadv.aav0693} deviation from the ideal cubic perovskite $\lvert \Delta\tau \rvert > 0.22$ (Section 5 of SM).

We actually provide an approach that evaluates the rate-limiting step at different temperatures based solely on the ground-state hydrogen bond length. To test this criterion, we selected the BaZrO$_3$ and SrCeO$_3$ systems and performed approximately 100 ps machine-learning molecular dynamics(MLMD) simulations. According to the critical hydrogen bond criterion of Fig.~\ref{Fig3}(c), BaZrO$_3$ and SrCeO$_3$ exhibit transfer-limited and rotation-limited behavior at 500 K, respectively, whereas both become rotation-transfer equivalent at 1000 K. And the MD simulations (Fig.~\ref{Fig3}(d)) show that at 500 K in BaZrO$_3$, the proton undergoes 33 rotations and only 3 transfers over the entire trajectory. When the temperature increases to 1000 K the proton exhibits long-range diffusion of 97 rotations and 48 transfers. Although rotation is still more frequent, the rotation-to-transfer ratio is drastically reduced (from 11 at 500 K to $\sim$
2 at 1000 K). For SrCeO$_3$ at 500 K, the proton performs 23 transfers and 2 rotations. At 1000 K, the transfer-to-rotation ratio decreases to about 2 (62 transfers vs 35 rotations). These characters are consistent with the prediction of the critical hydrogen bond criterion.

{\textbf{The targeted strategies of optimizing proton conductivity.}} Once the rate-limiting step at each temperature of a proveskit material is identified using the critical hydrogen bond criterion in Fig.~\ref{Fig3}(c), targeted strategies for optimizing proton conduction can be devised. Our results  and previous works\cite{rfs5-bnwt,PhysRevB.76.054307,https://doi.org/10.1111/jace.14839,draber2020nanoscale,HE2022111007} both show that A-site doping produces a more pronounced reduction in the rotation barrier, whereas B-site doping more effectively lowers the transfer barrier.  As shown in Fig.~\ref{Fig4}, low-valence A-site doping reduces the rotation barrier while increasing the transfer barrier, with the reduction in rotation barrier being more significant. In contrast, low-valence B-site doping lowers the transfer barrier but raises the rotation barrier, with a greater reduction observed in the transfer barrier. Overall, both A-site and B-site doping can effectively lower the energy barriers. Accordingly, when proton rotation is rate-limiting, doping with low-valence (e.g., +1) ions on A-site could elongate the initial hydrogen bond, lowering the rotation barrier. Conversely, when proton transfer limits the diffusion, lower-valence B-site dopants (e.g., +3) can shorten the initial hydrogen bond length, reducing the transfer barrier. By lowering the energy barrier of the rate-limiting step, the overall proton conductivity can be effectively enhanced.

\begin{figure}[htbp]
\centering
\includegraphics[scale=0.31]{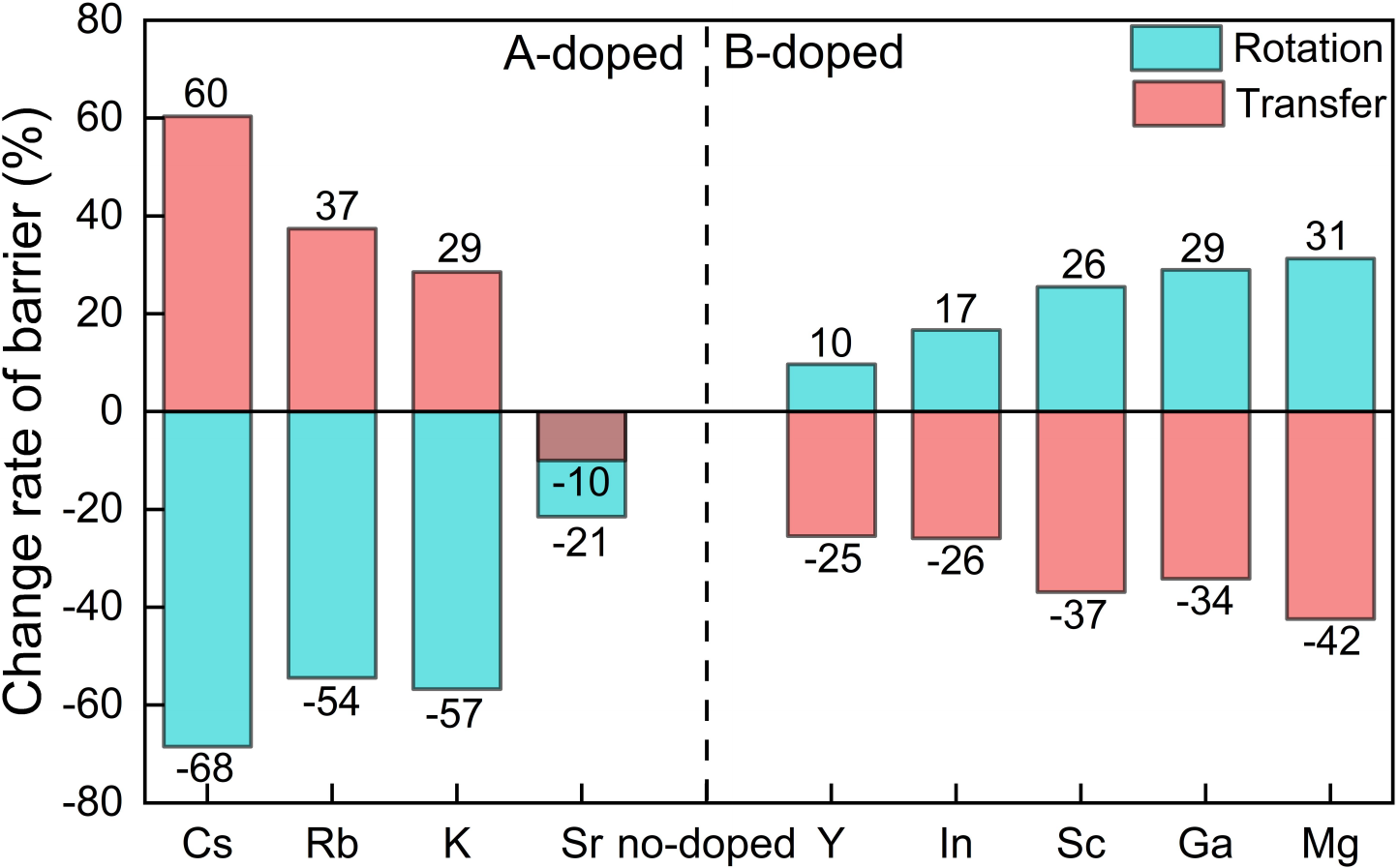}
\caption{\label{Fig4} Change rates of rotation and transfer barriers in different A-site and B-site doped systems relative to the undoped system.}
\end{figure}

Moreover, identifying the rate-limiting step is beneficial for screening rational materials for experimental research. Based on approximately 100~ps MD simulations of BaZrO$_3$ at 500 K and 1000 K, as well as SrCeO$_3$ at 500 K and 1000 K, we find that systems in which rotation and transfer are equivalent facilitate much more effective long-range proton diffusion compared with rate-limiting cases, consistent with previous studies\cite{doi:10.1021/acs.jpcc.0c04594,BJORKETUN20053035}. As shown in Fig.~\ref{Fig5}(a), at every time point the proton displacement is larger in the rotation-transfer equivalent systems (1000K BaZrO$_3$, 1000K SrCeO$_3$). Furthermore, by fitting the mean square displacement (MSD) curves for the four cases (Fig.~\ref{Fig5}(b)), we obtain diffusion coefficients that are about one order of magnitude higher than those of the rate-limiting systems. This difference goes beyond the enhancement expected from temperature effects alone, indicating that the distinct dynamical characteristics of rotation and transfer play a decisive role in determining the diffusivity. For example, a machine-learning study by Priya and Aluru \cite{priya2021accelerated} predicts a conductivity of $1.37\times10^{-7}$ S/cm for BaZrO$_3$, whereas LaScO$_3$ exhibits a much higher value of $2.82\times10^{-5}$ S/cm. This trend is consistent with our prediction in Fig. 3(b), where LaScO$_3$ falls into the rotation-transfer comparable regime with hydrogen-bond lengths below 2.05 \AA, while BaZrO$_3$ remains in the transfer-limited regime with longer hydrogen bonds. In addition, recent experimental and theoretical studies have shown that Sc doping in cubic Ba-based perovskites significantly enhances proton conductivity\cite{https://doi.org/10.1002/aenm.202000213,tsujikawa2025mitigating}. As shown in Fig. 3(a,b), pristine BaHfO$_3$ and BaZrO$_3$ exhibit hydrogen-bond lengths larger than 2.05 \AA, placing them in the transfer-limited regime. Upon B-site Sc doping, the hydrogen-bond lengths decrease to below 2.05 \AA, shifting the systems into the rotation-transfer comparable regime. Such a more balanced rotation-transfer barrier landscape is likely a contributing factor to the enhanced diffusion coefficients observed experimentally upon Sc doping.

\begin{figure}[htbp]
\centering
\includegraphics[scale=0.5]{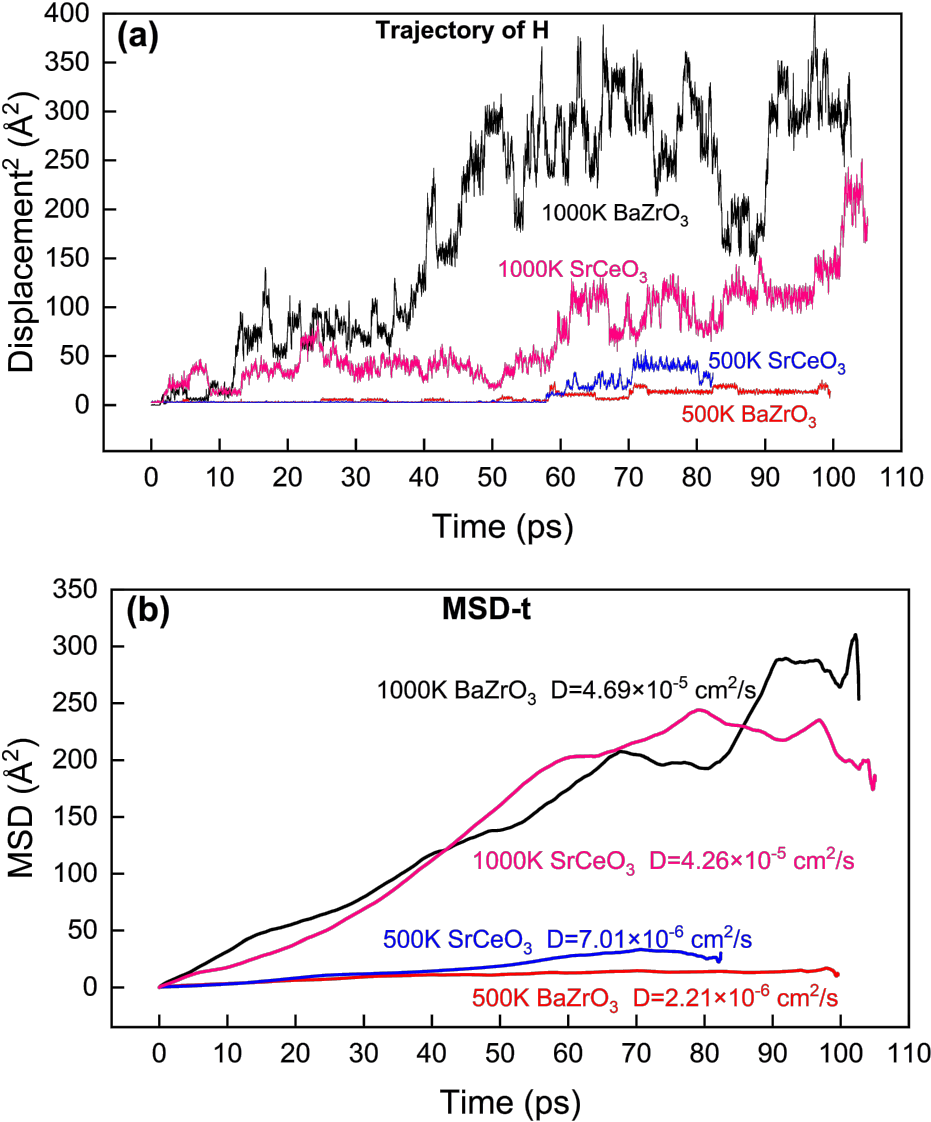}
\caption{\label{Fig5} (a) Time evolution of the squared proton displacement for the four systems. (b) Mean-square displacement (MSD) extracted from the 100~ps proton trajectories. The diffusion coefficients $D$ are obtained from linear fits to the MSD-t curves. The diffusion coefficients calculated here are consistent with experimental data and previous MD simulations reported in the literature\cite{kreuer2003proton,raiteri2011reactive}.}
\end{figure}

\noindent
\section{4. CONCLUSIONS}
Through a stepwise decomposition of the rotation and transfer energy barriers in BaHfO$_3$ and its A/B-site doped variants, we quantitatively demonstrate that the hydrogen bond length is the dominant factor controlling the barrier difference between the two processes through the hydrogen bond mediated lattice distortion. Using the residence time $t$ as a bridge connecting ground-state barrier differences to finite-temperature behavior, we extract critical hydrogen bond length criterions that identifies the rate-limiting step at different temperatures. Taking the 500 K criterion as an upper limit, we determine the hydrogen bond length (2.05 \AA) and the corresponding structural conditions under which rotation and transfer proceed at comparable rates. Furthermore, our proposal of targeted strategies to selectively lower the barrier of the identified rate-limiting step offers a rational pathway for optimizing proton conductivity in perovskite oxides.

\section*{Supporting Information}

Supporting Information. Details of climbing-image nudged elastic band (CI-NEB) calculations; energy barrier calculations for the elementary steps of proton rotation and transfer; net attractive effects of A-site and B-site dopants on protons; temperature-dependent distribution of the rotation-transfer equivalence region; calculations for perovskite systems with different structural features; schematics of proton rotation and transfer processes in A-site and B-site doped BaHfO$_3$.

\section*{AUTHOR INFORMATION}
\subsection*{Corresponding Authors}
\textbf{Jiajun Linghu} --- Department of Applied Physics, Chang'an University, Xi'an 710064, China; \\
Email: \href{mailto:linghujiajun@chd.edu.cn}{linghujiajun@chd.edu.cn}\\
\textbf{Ying Liang} --- College of Physics, Hebei Normal University, Shijiazhuang 050024, China; School of Physics and Astronomy, and Key Laboratory of Multiscale Spin Physics (Ministry of Education), Beijing Normal University, Beijing 100875, China \\
Email: \href{mailto:liang@bnu.edu.cn}{liang@bnu.edu.cn}\\
\textbf{Kunde Yang} --- Ocean Institute of Northwestern Polytechnical University,Taicang, Jiangsu  215400, China; \\
Email: \href{mailto:ykdzym@nwpu.edu.cn}{ykdzym@nwpu.edu.cn}\\
\textbf{Tianxing Ma} --- School of Physics and Astronomy, and Key Laboratory of Multiscale Spin Physics (Ministry of Education), Beijing Normal University, Beijing 100875, China; \\
Email: \href{mailto:txma@bnu.edu.cn}{txma@bnu.edu.cn}\\
\textbf{Zhi-Peng Li} --- State Key Laboratory of Flexible Electronics (LOFE) \& Institute of Flexible Electronics (IFE), Northwestern Polytechnical University, 127 West Youyi Road, Xi'an, 710072, China; \\
Email: \href{mailto:iamzpli@nwpu.edu.cn}{iamzpli@nwpu.edu.cn}

\subsection*{Authors}
\textbf{Hang Ma} --- School of Physics and Astronomy, and Key Laboratory of Multiscale Spin Physics (Ministry of Education), Beijing Normal University, Beijing 100875, China\\
\textbf{Nannan Han} --- State Key Laboratory of Flexible Electronics (LOFE) \& Institute of Flexible Electronics (IFE), Northwestern Polytechnical University, 127 West Youyi Road, Xi'an, 710072, China\\
\textbf{Peng Feng} --- State Key Laboratory of Flexible Electronics (LOFE) \& Institute of Flexible Electronics (IFE), Northwestern Polytechnical University, 127 West Youyi Road, Xi'an, 710072, China\\
\textbf{Yuling Zhuo} --- State Key Laboratory of Flexible Electronics (LOFE) \& Institute of Flexible Electronics (IFE), Northwestern Polytechnical University, 127 West Youyi Road, Xi'an, 710072, China\\

\subsection*{Author contributions}

Y.L., K.Y., T.M., and Z.-P.L. directed this project. H.M. and J.L. performed the computational simulations. All authors discussed the results and contributed to the manuscript.

\subsection*{Notes}

The authors declare no competing interests.

\section*{ACKNOWLEDGMENT} 
This work was supported by National Natural Science Foundation of China (Nos. 12404463 and 12474218), and Beijing Natural Science
Foundation (Nos. 1242022 and 1252022),  the Double First-Class Construction Fund for Teacher Development Projects (0515024GH0201201 and 0515024SH0201201), and the Key Research and Development Program of Shaanxi (2024GX-YBXM-456 and 2024GX-YBXM-565).

\bibliography{reference}




\newpage
\noindent
\clearpage
\setcounter{figure}{0}\setcounter{table}{0}\setcounter{equation}{0}
\renewcommand{\thefigure}{S\arabic{figure}}
\renewcommand{\thetable}{S\arabic{table}}
\renewcommand{\theequation}{S\arabic{equation}}
\onecolumngrid

\begin{center}{\large\bfseries Supporting Information}\end{center}

\noindent
\section{S1. Setting of the CI-NEB calculations}

\begin{figure*}[htbp]
\centering
\includegraphics[scale=0.5]{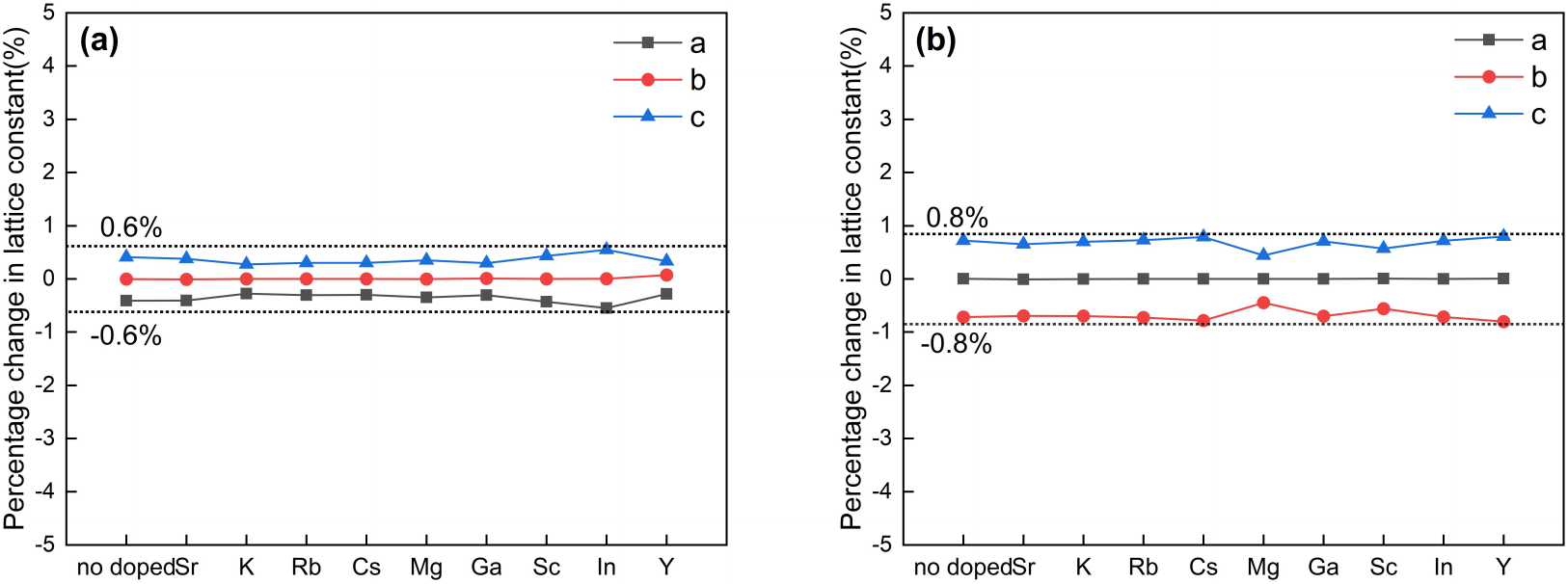}
\caption{\label{FigS1} (a) Percentage change in lattice constants along the a, b, and c directions before and after proton rotation. (b) Percentage change in lattice constants along the a, b, and c directions before and after proton transfer. }
\end{figure*}

Figure~\ref{FigS1} validates the use of fixed-supercell (ISIF=2) calculations for proton rotation and transfer energy barriers. We performed full relaxations (ISIF=3) for the initial and final states of proton rotation and transfer in all A- and B-site doped BaHfO$_3$ systems. The resulting changes in lattice constants along the a, b, and c directions are less than 0.6\% for rotation steps and less than 0.8\% for transfer steps. These minimal variations indicate that proton motion has negligible impact on lattice dimensions, justifying the use of fixed-supercell CI-NEB to obtain reliable energy barriers.

\noindent
\section{S2. Calculation of energy barriers for the elementary steps of proton rotation and transfer}

\begin{figure*}[htbp]
\centering
\includegraphics[scale=0.47]{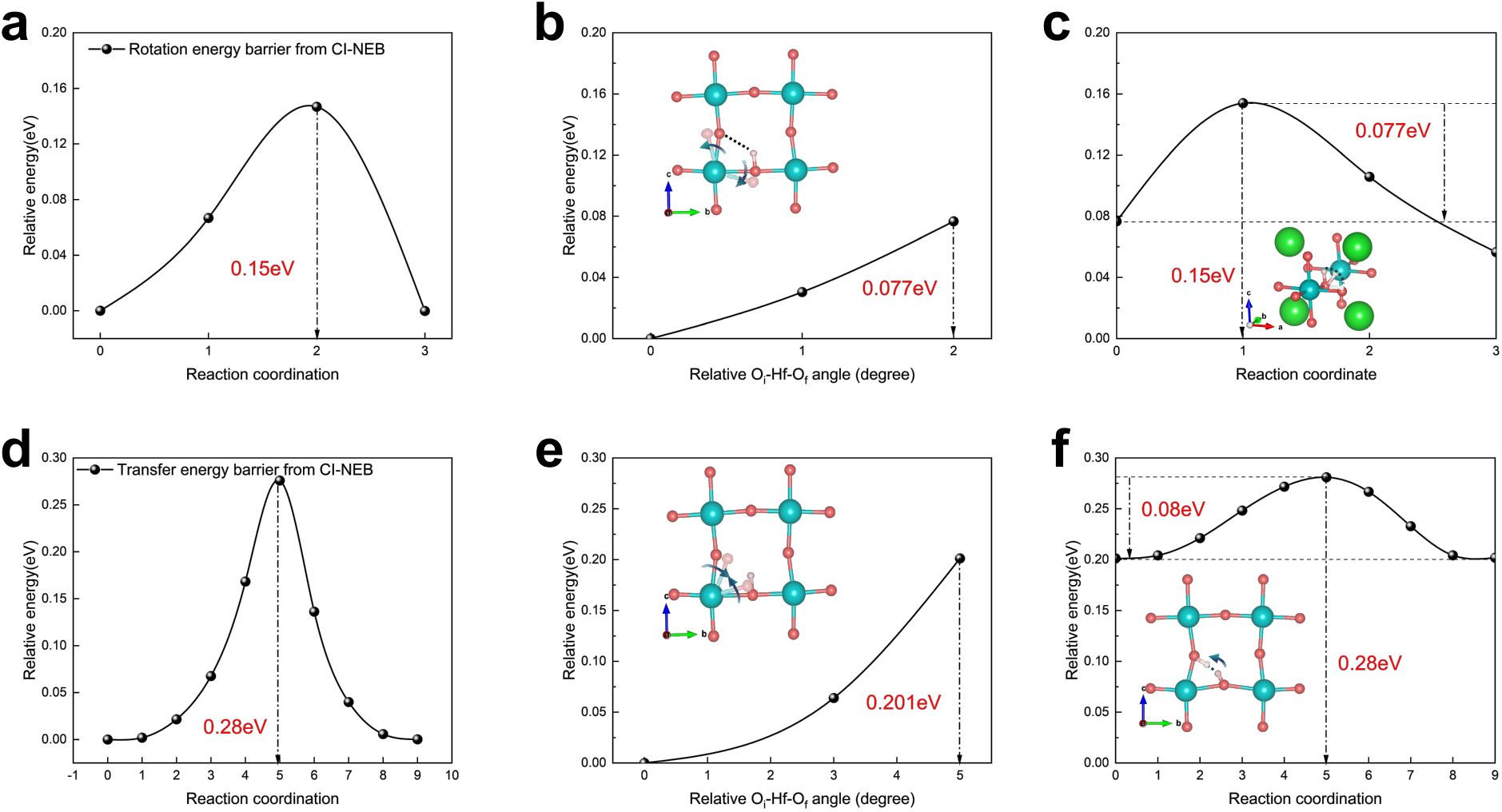}
\caption{the energy barriers for the elementary steps underlying proton rotation and transfer in the BaHfO$_3$. (a) the rotation energy barrier obtained from CI-NEB calculation. (b) the energy barriers for the outward bending of O$_i$-B-O$_f$ motion during the proton rotation. (c) the energy barriers for the reorientation of proton during the rotation. (d) the transfer energy barrier obtained from CI-NEB calculation. (e) the energy barriers for the inward bending of O$_i$-B-O$_f$ motion during the proton transfer. (f) the energy barriers for the proton jumping during the proton transfer. } 
\label{FigS2}
\end{figure*}

Figure~\ref{FigS2} shows the comparison between the calculated energy barriers using CI-NEB and the energy barriers based on the elementary steps underlying rotation and transition for the BaHfO$_3$ system. During the proton rotation, the energy barriers for the outward O$_i$-B-O$_f$  bending motion and the reorientation of proton are 0.077 eV (Fig.~\ref{FigS2}(b)) and 0.077 eV (Fig.~\ref{FigS2}(c)), respectively. Therefore, the total energy barrier based on these two elementary steps is 0.15 eV , which is consistent with the value of 0.15 eV obtained from our CI-NEB calculations(Fig.~\ref{FigS2}(a)). During the proton transfer, the energy barriers for the inward O$_i$-B-O$_f$ bending motion and the proton jumping are calculated to be 0.20 eV and 0.08 eV, respectively. The total energy barrier based on these two elementary steps is 0.28 eV, which is also consistent with the CI-NEB calculation result of 0.28 eV (Fig.~\ref{FigS2}(d)). These results confirm the validity of the proton rotation and transfer mechanism based on these elementary steps.

\noindent
\section{S3. The net attractive effect of different A-site and B-site dopants on the proton}

\begin{figure}[htbp]
\centering
\includegraphics[scale=0.37]{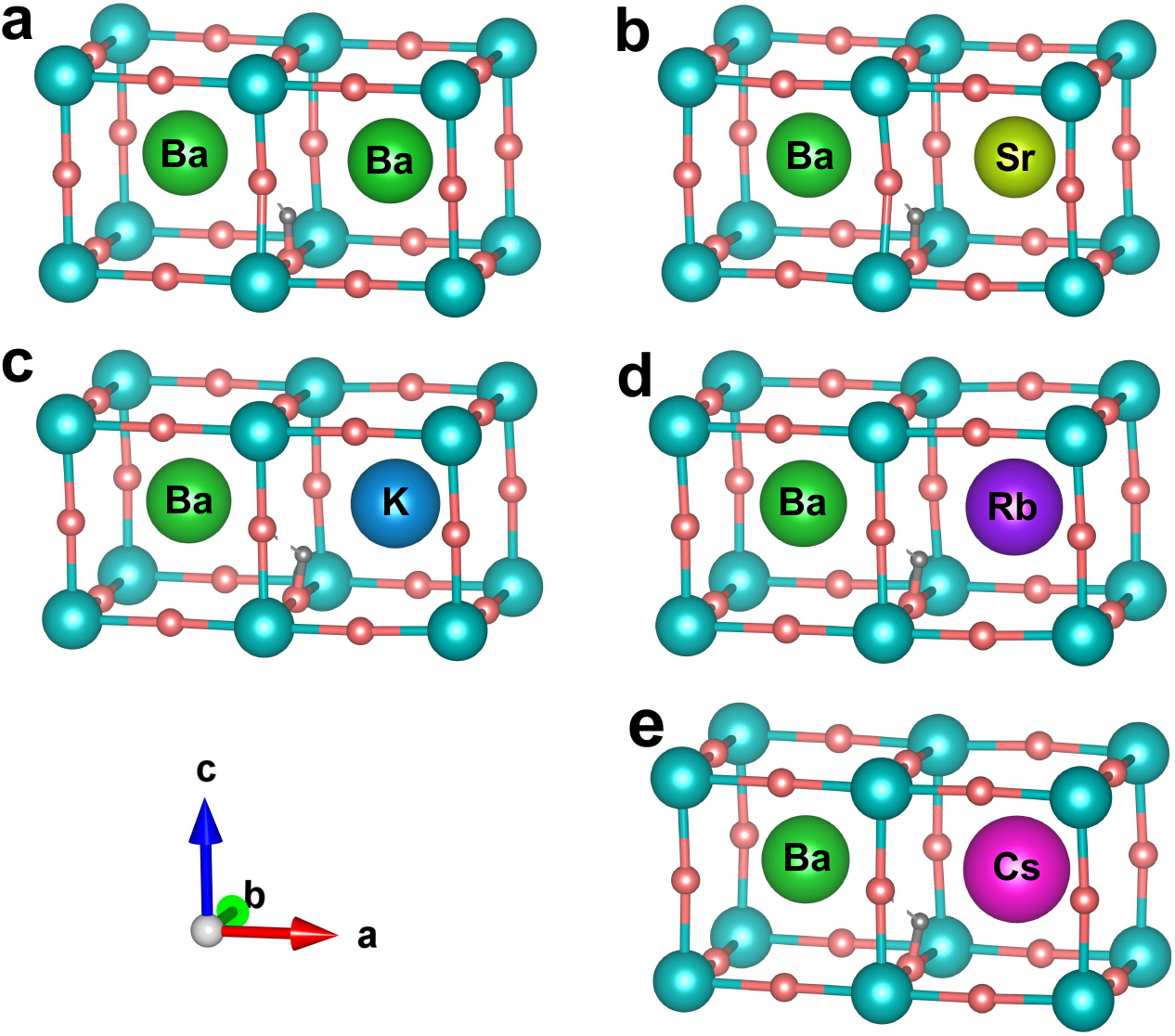}
\caption{illustrative diagram of the proton orientation in the BaHfO$_3$ and different A-site doped systems. (a) undoped, (b) Sr-doped, (c) K-doped, (d) Rb-doped, (e) Cs-doped. } 
\label{FigS5}
\end{figure}

Figure~\ref{FigS5} displays the illustrative diagram of the proton orientation in the BaHfO$_3$ and different A-site doped systems. For systems doped with monovalent dopant ions such as K$^{+}$, Rb$^{+}$,and Cs$^{+}$, the proton tilts towards the unit cell space occupied by the A-site dopant ions due to the net attractive force, as shown in Fig.~\ref{FigS5}(c)-(e), resulting in the formation of relatively long hydrogen bonds(TABLE I). Furthermore, due to the larger ionic radii of K$^{+}$, Rb$^{+}$,and Cs$^{+}$(Ba$^{2+}$= 1.61 \AA, K$^{+}$ = 1.64 \AA, Rb$^{+}$ = 1.72 \AA, Cs$^{+}$ = 1.88 \AA\cite{Shannon:a12967}), the lattice of the unit cell expands, which also leads to an increase in hydrogen bond length. In previous studies, we found that as the ionic radius of the A-site dopant increases, the average M$_a$-O bond length in the corresponding unit cell also increases, indicating a more pronounced lattice expansion effect. Consequently, the strength of the hydrogen bonds formed in the system decreases progressively in the order of K$^{+}$, Rb$^{+}$,and Cs$^{+}$ doping (TABLE I). For systems doped with the divalent ion Sr$^{2+}$, the unit cell contracted due to the smaller ionic radius of Sr$^{2+}$ (Fig.~\ref{FigS5}(b)). This effect shortens the hydrogen bond length formed between the proton and the acceptor oxygen ion theoretically.

\begin{figure}[htbp]
\centering
\includegraphics[scale=0.37]{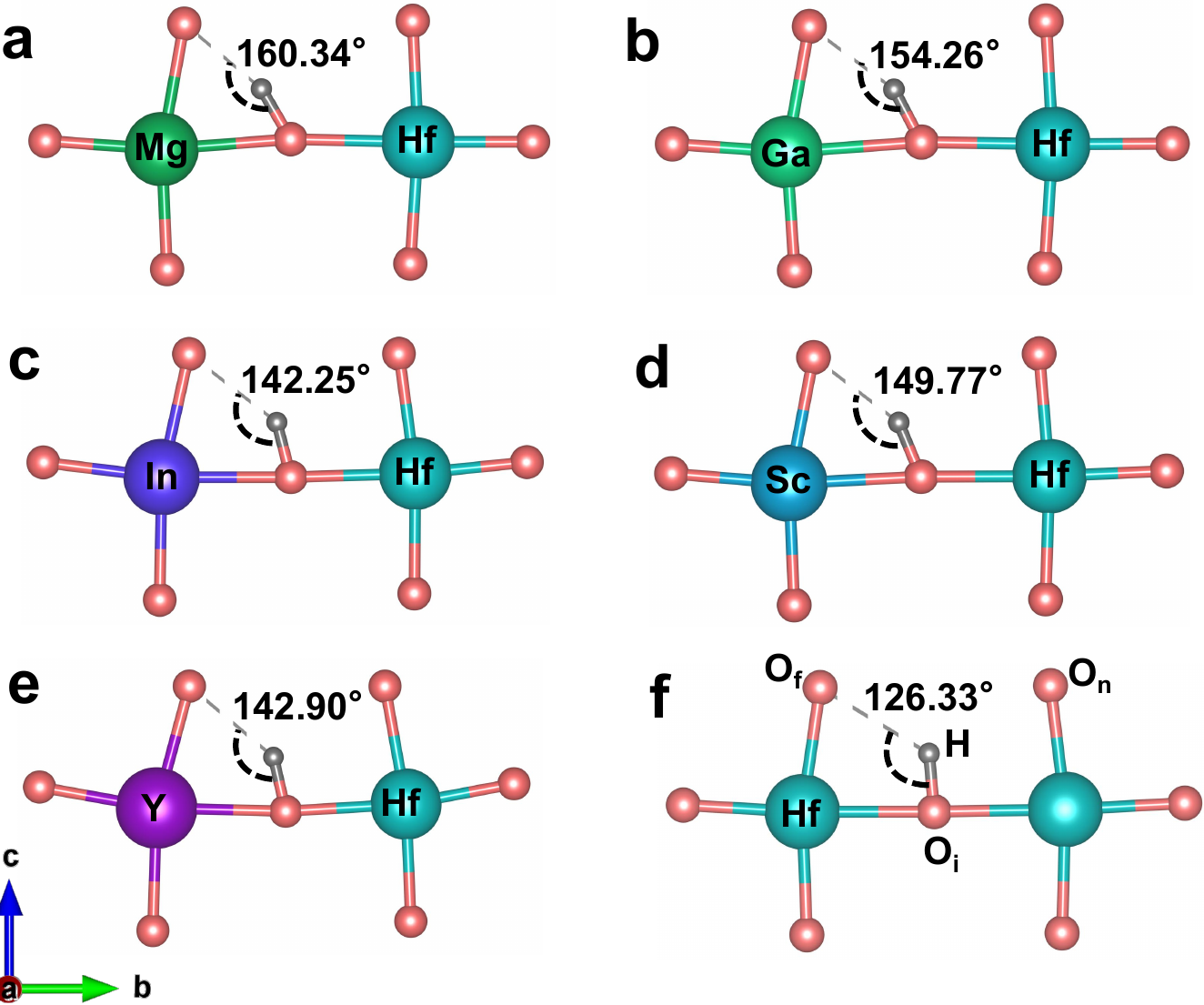}
\caption{illustrative diagram of the net attractive force on the proton exerted by B-site doped ions. (a) Mg-doped, (b) Ga-doped, (c) Sc-doped, (d) In-doped, (e) Y-doped, (f) undoped } 
\label{FigS6}
\end{figure}

\begin{table}[htbp]
\centering
\caption{Ionic radii of different B-site dopant ions\cite{Shannon:a12967} and the corresponding octahedral unit cell parameters }
\label{TabS1}
\renewcommand{\arraystretch}{1.5}
\begin{tabular}{|c|c|c|}
\hline
\textbf{Dopant} & \textbf{Ionic radius(\AA)} & \textbf{The average of M$_b$-O(\AA)} \\ \hline
Ga       & 0.62  & 2.09   \\
Hf       & 0.71  & 2.10   \\
Mg       & 0.72  & 2.13   \\
Sc       & 0.75  & 2.13   \\
In       & 0.8   & 2.17   \\
Y        & 0.9   & 2.22   \\ \hline
\end{tabular} 
\end{table}

Figure~\ref{FigS6} shows the illustrative diagram of the net attractive force on the proton exerted by B-site doped ions. Since Mg$^{2+}$ has the lowest valence, it exerts the strongest net attractive force on the proton, resulting in the proton tilting to the greatest extent and the O$_i$-H...O$_f$ hydrogen bond approaching the most linear configuration (close to 180$^\circ$)(Fig.~\ref{FigS6}(a)) , which the formed hydrogen bond H...O$_f$ is the shortest (TABLE I). For the trivalent dopant ions  Ga$^{3+}$, In$^{3+}$, Sc$^{3+}$,and Y$^{3+}$, it can be observed that as the ionic radius increases, the formed hydrogen bonds weaken (TABLE I). This is because a larger ionic radius such as In$^{3+}$ and Y$^{3+}$ leads to the expansion of the surrounding octahedron (TAB.~\ref{TabS1}. By measuring the average distance between the B-site dopant ion and the six coordinated oxygen ions.), which shortens the distance between the proton and the oxygen ion O$_n$ in the adjacent octahedron, promoting the formation of the hydrogen bond H...O$_n$. However, the attraction between the proton and the oxygen ion O$_n$ in the adjacent octahedron weakens the net attractive force of the B-site dopant ions on the proton, thus reducing the tilt of the proton. As a result, the degree of  decreases(Fig.~\ref{FigS6}(d),(e)), leading to a longer hydrogen bond length(TABLE I)\cite{PhysRevB.76.054307,C3TA12870A}.

\noindent
\section{S4. Temperature dependence of the distribution of the rotation-transfer equivalence region}

\begin{figure}[htbp]
\centering
\includegraphics[scale=0.4]{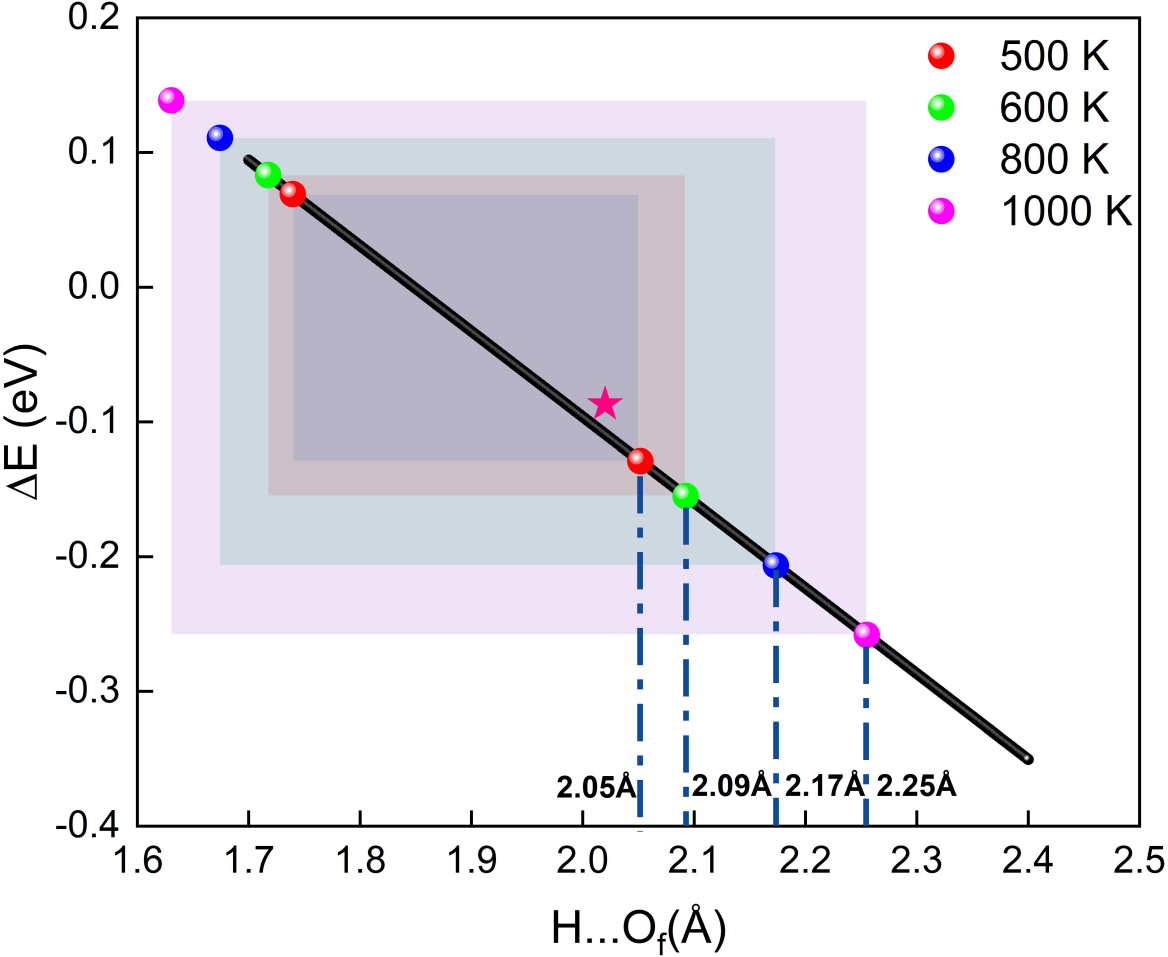}
\caption{\label{FigS7} The positions of the critical hydrogen bond points at 500~K, 600~K, 800~K, and 1100~K are shown, where the lower-right data correspond to the critical points for the transfer-limiting, and the upper-left points are the rotation-limiting.}
\end{figure}

As shown in Fig. \ref{FigS7}, the intermediate region where rotation and transfer are equivalent is smallest at 500~K and expands progressively with increasing temperature, always encompassing the region at the previous temperature. This reflects the higher sensitivity of the residence time $\tau$ to energy at lower temperatures. Therefore, the 500~K result effectively establishes an upper-bound criterion: if a material exhibits a hydrogen bond length shorter than 2.05~\AA (the star symbols in Fig. \ref{FigS7}) and thus lies in the rotation-relevant region at 500~K, then rotation must also be considered at practical operating temperatures (300–800~\si{\celsius}), where this region is even broader.

\noindent
\section{S5. Calculations on perovskite systems with different structural features}

\begin{table}[htbp]
\centering
\caption{the rotation and transfer energy barriers(E$_{rot}$,E$_{trans}$), along with the corresponding hydrogen bond lengths(H...O$_f$), calculated via CI-NEB for perovskite systems with distinct structural characteristics. }
\label{TabS2}
\renewcommand{\arraystretch}{1.5}
\begin{tabular}{|c|c|c|c|}
\hline
\textbf{} & \textbf{E$_{rot}$(eV)} & \textbf{E$_{trans}$(eV)} & \textbf{H...O$_f$(\AA)} \\ \hline
BaNbO$_3$ & 0.13 & 0.33 & 2.30 \\
BaTiO$_3$ & 0.15 & 0.24 & 2.22 \\
BaZrO$_3$ & 0.15 & 0.27 & 2.19 \\
BaHfO$_3$ & 0.15 & 0.28 & 2.15 \\
SrTiO$_3$ & 0.18 & 0.27 & 2.02 \\
SrNiO$_3$ & 0.13 & 0.12 & 1.99 \\
BaZr$_x$La$_{1-x}$O$_3$ & 0.08 & 0.29 & 2.09 \\
BaZr$_x$Gd$_{1-x}$O$_3$ & 0.11 & 0.21 & 2.02 \\
BaZr$_x$Y$_{1-x}$O$_3$ & 0.12 & 0.22 & 2.01 \\
BaZr$_x$Sc$_{1-x}$O$_3$ & 0.16 & 0.20 & 1.92 \\
BaZr$_x$Ga$_{1-x}$O$_3$ & 0.17 & 0.19 & 1.83 \\
SrNbO$_3$ & 0.03 & 0.25 & 2.09 \\
BaCeO$_3$ & 0.56 & 0.80 & 2.06 \\
SrZrO$_3$ & 0.22 & 0.36 & 2.03 \\
LaScO$_3$ & 0.02 & 0.14 & 1.97 \\
SrCeO$_3$ & 0.27 & 0.14 & 1.71 \\ \hline
\end{tabular} 
\end{table}

Table~\ref{TabS2} lists the calculated rotation and transfer energy barriers, along with the corresponding hydrogen bond lengths for perovskite systems with distinct characteristics(cubic structures, B-site doped systems, and orthorhombic structures). These results serve to assess the generality of the hydrogen bond as the dominant factor and to validate the critical hydrogen bond length identified in Fig. 3. In the orthorhombic structures considered here, octahedral tilting leads to inequivalent sites, resulting in asymmetric barriers between the initial and final states (with energy differences exceeding 0.1 eV). To address this, for both rotation and transfer processes we consistently take the higher-energy state as the initial state and the lower-energy state as the final state. In evaluating the rotation-transfer barrier difference, this asymmetry is effectively canceled, yielding a more intrinsic measure of the barrier difference. Taking SrCeO$_3$ as an example, the rotation-transfer barrier difference is 0.14 eV with the kinetic resolved activation (KRA) correction (where the barrier are referenced to the average energy of the initial and final states) and 0.13 eV without it.

Based on the distribution of these systems with distinct characteristics in the H...O$_f$-$\Delta E$ diagram (Fig.~3(c)), we identify the structural conditions under which proton rotation becomes competitive with transfer. Cubic perovskites typically exhibit longer hydrogen bonds and higher transfer barriers due to their symmetric structures. However, compounds such as SrNiO$_3$ and SrTiO$_3$ deviate from this trend, as their smaller A-site cation radius reduces the lattice constant, shortening hydrogen bonds ($<$ 2.05~\AA) and making rotation non-negligible. As shown in Fig.~3(c), SrTiO$_3$, with a lattice constant of 3.95~\AA, lies near the threshold, whereas Ba-based systems (e.g., BaHfO$_3$, BaZrO$_3$) have larger lattice constants and remain transfer-limited. Thus, cubic perovskites with lattice constants $\leq$ 3.95~\AA require inclusion of rotational contributions.; In low-valent B-site doped systems, dopants tend to attract protons electrostatically, promoting shorter hydrogen bonds and lowering transfer barriers. However, La-doped BaZrO$_3$ remains transfer-limited due to the large ionic radius ($\Delta r$ = 0.312~\AA), which induces local octahedral expansion and elongates hydrogen bonds ($>$ 2.05~\AA, Fig. S4). In contrast, smaller-radius dopants (Gd$^{3+}$, Y$^{3+}$, Sc$^{3+}$, Ga$^{3+}$) promote rotation-dominated behavior. Consequently, In the low-valent B-site doped systems with larger ionic radius mismatches $\Delta r < 0.312$~\AA, rotation should be considered; For orthorhombic perovskites, mismatch between A-site and B-site cation sizes induces octahedral tilting and short hydrogen bonds. This distortion is quantified by Bartel's tolerance factor\cite{doi:10.1126/sciadv.aav0693}, where larger $\lvert \Delta\tau \rvert$ (deviation from the ideal ($\tau$= 3.5)) indicates stronger distortion. SrZrO$_3$, with $\lvert \Delta\tau \rvert$ = 0.22, lies near the critical point. Materials with $\lvert \Delta\tau \rvert < 0.22$ (e.g., BaCeO$_3$, SrNbO$_3$) are transfer-limited, while those with $\lvert \Delta\tau \rvert > 0.22$ (e.g., SrCeO$_3$, LaScO$_3$) show rotation-limited or transfer-rotation comparable behavior. Therefore,the orthorhombic perovskites with $\lvert \Delta\tau \rvert > 0.22$ should include rotational pathways in describing proton transport.

\noindent
\section{S6. Schematics of proton rotation and transfer processes in A-site and B-site doped BaHfO$_3$}

\begin{figure}[htbp]
\centering
\includegraphics[scale=0.35]{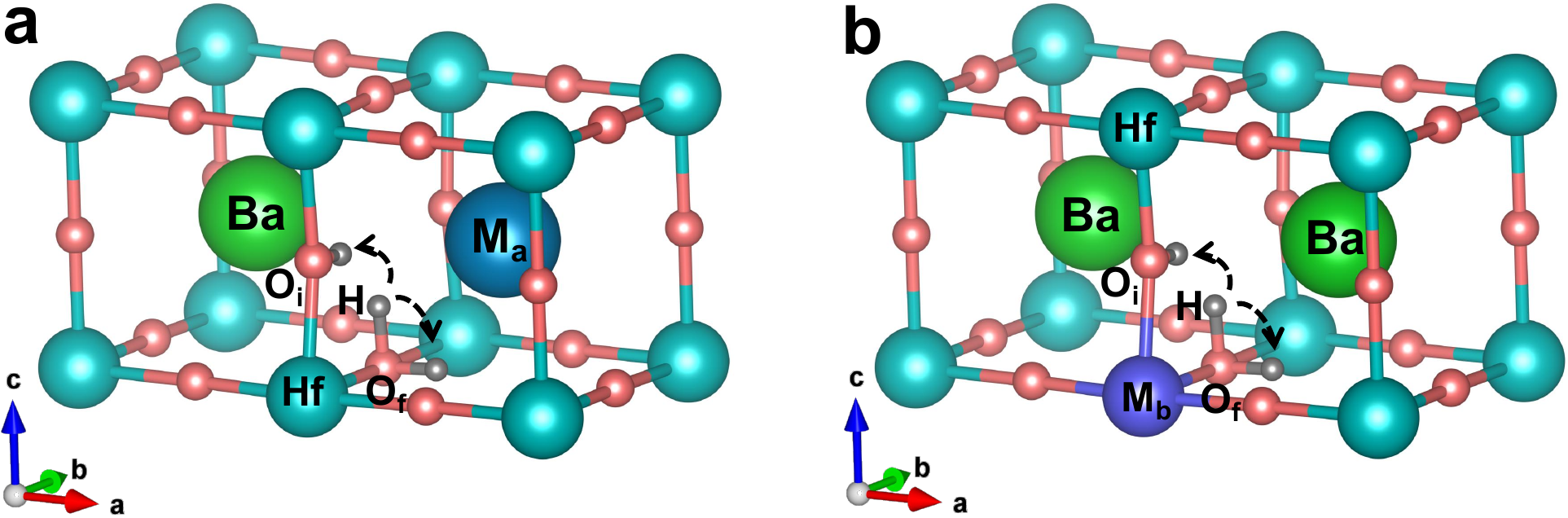}
\caption{(a) Calculated rotation and transfer processes for A-site doped (M$_a$) BaHfO$_3$ (b) Calculated rotation and transfer processes for B-site doped (M$_b$) BaHfO$_3$ .} 
\label{FigS8}
\end{figure}

Figure~\ref{FigS8} displays the calculated rotation and transfer processes for A-site (M$_a$) and B-site doped (M$_b$) BaHfO$_3$. The doping concentrations of  A-site (M$_a$ = Sr$^{2+}$, K$^{+}$, Rb$^{+}$, Cs$^{+}$) and B-site (Mb = Mg$^{2+}$, Ga$^{3+}$, In$^{3+}$, Sc$^{3+}$, Y$^{3+}$) are 0.125, based on a $2 \times 2 \times 2$ supercell. For the A-site doping system, we computed energy barrier of the proton rotation process around the A-site dopant ion (FIG.~\ref{FigS1}(a)), as well as the process of intra-octahedral proton transfer in the unit cell containing the A-site dopant ion (Fig.~\ref{FigS8}(a), proton transfer from O$_i$ to O$_f$). For the B-site doping system, we computed energy barrier of the intra-octahedral proton transfer process around the B-site dopant ion (Fig.~\ref{FigS8}(b), proton transfer from O$_i$ to O$_f$), as well as the proton rotation process at a specific oxygen site within the octahedron of the B-site dopant ion (FIG.~\ref{FigS1}(b)). These calculations are performed to investigate the impact of A-site and B-site doping on the proton rotation and transfer.

\end{document}